\PassOptionsToPackage{fontset=none}{ctex}
\documentclass{mosi}
\usepackage{array,tabularx,needspace,flafter}
\usepackage{tikz}
\usetikzlibrary{arrows.meta,positioning}
\renewcommand{\arraystretch}{1.12}
\setCJKmonofont[Extension=.ttf,Path=fonts/]{NotoSansSC-Prompts.ttf}
\newenvironment{CJK}[2]{}{}

\newcommand{\modelname}{Soloist}
\newcommand{\Khat}{\widehat K}
\newcommand{\Kmax}{K_{\mathrm{max}}}

\newcommand{\Rjudge}{R_{\mathrm{judge}}}

\DeclareMathOperator{\clip}{clip}
\newcolumntype{Y}{>{\raggedright\arraybackslash}X}
\newcolumntype{L}[1]{>{\raggedright\arraybackslash}p{#1}}

\newenvironment{promptbox}{\begin{tcolorbox}[breakable,colback=gray!4,colframe=gray!45,boxrule=0.4pt,arc=0pt,left=2mm,right=2mm,top=1mm,bottom=1mm]\small\ttfamily\raggedright}{\end{tcolorbox}}

\pretocmd{\section}{\Needspace{8\baselineskip}}{}{}
\pretocmd{\subsection}{\Needspace{5\baselineskip}}{}{}
\title{Soloist: A Generative Model for\\Blind Voice Separation in the Wild}
\renewcommand{\authorlist}{{\fontsize{12}{17}\selectfont\bfseries
Chen Yang\textsuperscript{1,2,*}\quad ChuFan Yu\textsuperscript{1,*}\quad Botian Jiang\textsuperscript{1,2}\quad Jingqi Chen\textsuperscript{1}\\
Wenxuan Wang\textsuperscript{1,2}\quad Yitian Gong\textsuperscript{1}\quad Zhaoye Fei\textsuperscript{1}\quad Xipeng Qiu\textsuperscript{1,2,\textdagger}\par}
\vspace{2mm}
{\fontsize{10}{13}\selectfont
\textsuperscript{1}Fudan University\qquad\textsuperscript{2}Shanghai Innovation Institute\par}
\vspace{1mm}
{\fontsize{9}{12}\selectfont\textsuperscript{*}Equal contribution\qquad\textsuperscript{\textdagger}Corresponding author\par}
}
\hypersetup{pdftitle={Soloist Technical Report},pdfauthor={Chen Yang, ChuFan Yu, Botian Jiang, Jingqi Chen, Wenxuan Wang, Yitian Gong, Zhaoye Fei, Xipeng Qiu}}
\abstract{Real-world voice separation must recover individual voices from overlapping speech, environmental noise, and music, yet isolated source recordings are rarely available for supervision. We introduce \modelname{}, a generative model for blind voice separation that autoregressively generates individual voice streams as audio-codec sequences conditioned on the mixture, without requiring a target-speaker query or a predefined voice count. We propose a two-stage training paradigm to learn from synthetic mixtures and adapt to real recordings. The first stage combines supervised separation on synthetic mixtures with isolated source targets and an auxiliary task that predicts speaker timelines from real recordings, learning who speaks when from natural acoustics. The second stage applies reinforcement learning to real recordings, with rewards supplied by a model judge that compares candidate separations based on voice purity and content preservation without isolated source targets. Across diverse scenarios, \modelname{} recovers clean speech from multi-speaker overlap and noisy recordings, and extracts singing vocals from music accompaniment while maintaining relatively high pitch accuracy and intelligibility. On real-world recordings with unknown voice counts, \modelname{} also separates voices well. Finally, we integrate \modelname{} into a full-duplex conversation processing pipeline that assembles separated outputs into continuous speaker tracks, demonstrating its practical utility in real conversations.
}
\begin{document}
\maketitle
\section{Introduction}\label{sec:intro}
People speak over one another in conversations, films, and online recordings. Recovering their individual voices requires a separator to identify how many voices are present, retain each voice's content and identity, and exclude unrelated sound. Training references are less complete: a natural recording often comes with a transcript or speaker timeline, but rarely with the isolated waveforms that produced the mixture.

Supervised separators learn this mapping from mixtures constructed with known sources~\citep{luo2019convtasnet,subakan2021sepformer}. Their targets are precise, while the acoustics and source combinations available for construction constrain the training distribution. Learning from unpaired mixtures has also been studied: mixture invariant training (MixIT) uses mixture-level reconstruction without isolated sources~\citep{wisdom2020mixit}, including adaptation to real meetings~\citep{sivaraman2021mixit}. Our focus is a complementary setting in which a generative separator can use several incomplete references and can be updated on natural audio through model-generated feedback. \Cref{app:related} positions this approach relative to prior work.

We introduce \modelname{}, a generative model for blind voice separation in real recordings. Given a single-channel mixture, the model generates one audio-codec sequence per voice without requiring a target-speaker query or a supplied voice count. Each sequence is decoded into an individual voice waveform.

\begin{figure}[htbp]
\centering
\includegraphics[width=\linewidth]{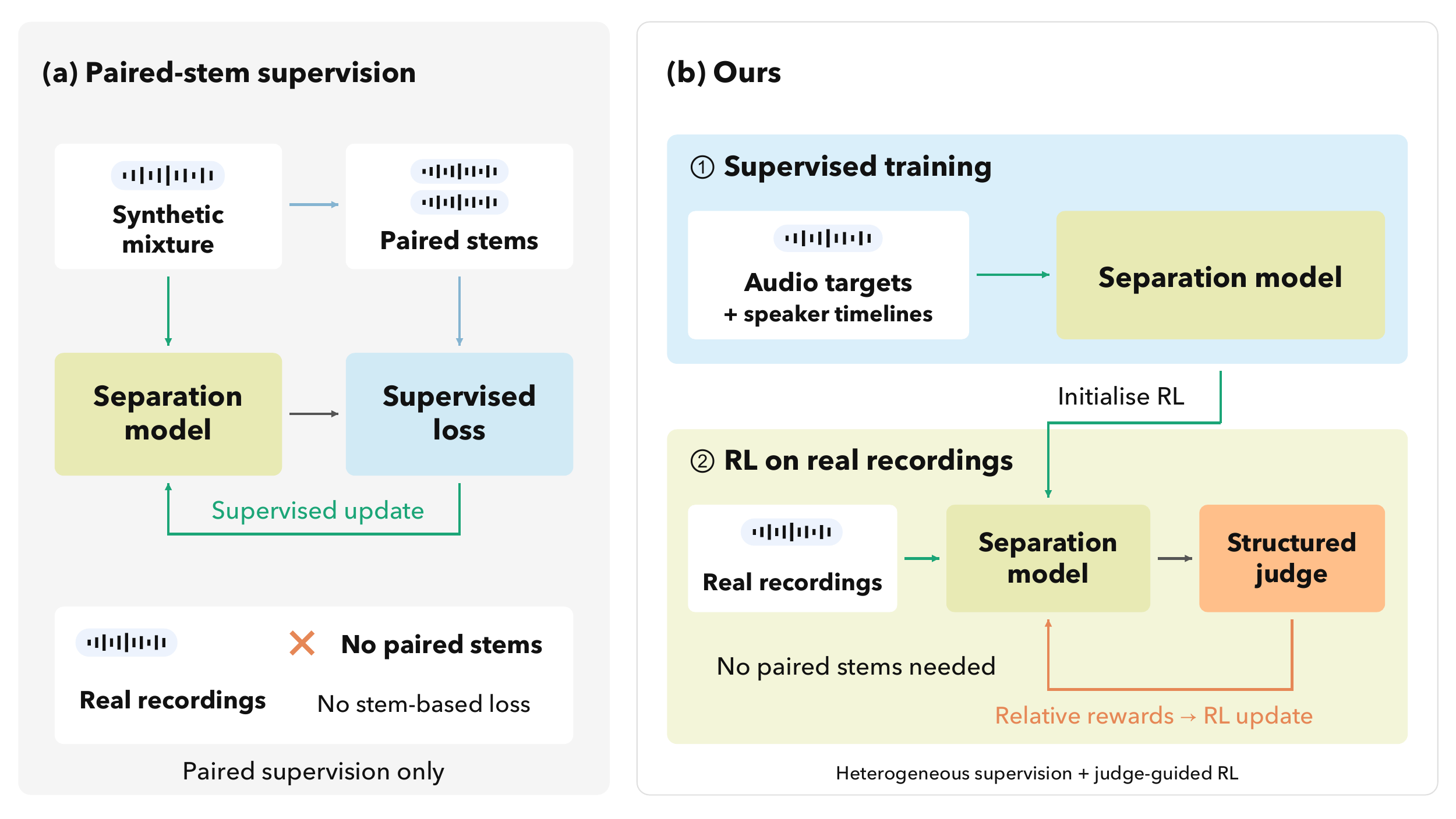}
\caption{Training with incomplete references. (a) Paired-stem supervision compares predicted audio with known sources; natural recordings usually lack these targets. (b) Our recipe combines audio targets and speaker timelines before adaptation. RL uses model-judge rewards on real recordings. Training details are given in \cref{app:training}.}
\label{fig:dependency}
\end{figure}

Our training recipe follows the references available for each recording (\cref{fig:dependency}). Constructed mixtures provide source targets; a fixed music separator provides pseudo-vocal targets for songs; and speaker timelines provide text targets for natural recordings without isolated waveforms. These examples share one supervised objective. After supervised training, reinforcement learning adapts the model on real recordings by sampling alternative separations of each mixture. A model judge supplies the reward: it describes tracks independently, compares the resulting reports, and checks consistency across candidate orders.

The judge must distinguish useful separation from fluent but incorrect resynthesis. Signal checks detect degenerate outputs and likely copies. The input mixture is included as an anonymous baseline, allowing the judge to compare each separation with returning the input unchanged. To check whether candidate order affects its preferences, the judge ranks the same candidates twice, first in a shuffled order and then in the reverse order. Groups with inconsistent rankings, or reports that cannot distinguish candidate quality, are excluded from RL updates. The remaining groups provide relative rewards that favor purer individual voices and more complete coverage of the audible content.

This report makes three contributions. First, we formulate blind voice separation as serial codec-sequence generation with an explicit voice count, and show that predicting speaker timelines on real recordings without isolated sources improves count inference. Second, we propose a model-judge reward for RL on real recordings without isolated sources: tracks are described independently, candidates are ranked in text against the unprocessed mixture as an anonymous anchor, and groups whose preferences change substantially when the candidate order is reversed are discarded. Third, we evaluate the model on synthetic, noisy, singing, and real multi-speaker recordings, and show how it is used in a full-duplex conversation processing pipeline.

\FloatBarrier
\section{Related Work}\label{app:related}

\subsection{Supervised Separation and Variable Source Counts}
Paired-source separation learns to recover isolated waveforms from constructed mixtures. Conv-TasNet operates in a learned time-domain representation~\citep{luo2019convtasnet}; SepFormer and MossFormer develop attention-based sequence models~\citep{subakan2021sepformer,zhao2023mossformer}; and TF-GridNet combines full-band and sub-band modeling in the time-frequency domain~\citep{wang2023tfgridnet}. These approaches establish strong reconstruction baselines, while the choice of training mixtures determines the acoustic conditions and source combinations available to supervision. LibriMix provides a reproducible construction for studying speech separation and generalization~\citep{cosentino2020librimix}.

Output assignment and source count are distinct problems. Permutation-invariant training handles the arbitrary ordering of reference sources~\citep{yu2017pit,kolbaek2017upit}, whereas recursive separation extracts speakers successively to accommodate unknown counts~\citep{takahashi2019recursive}. SR-CorrNet instead adapts the number of output streams with an attractor-based split module~\citep{shin2026srcorrnet}. \modelname{} represents count, stream boundaries, and voice audio in one serial response. Its supplied-count and blind evaluations therefore probe different capabilities: recovering voices under a known count and deciding how many streams to produce (\cref{sec:eval}).

\subsection{Neural Audio Codecs and Generative Separation}
SoundStream and EnCodec learn compressed audio representations that can be decoded back into waveforms~\citep{zeghidour2021soundstream,defossez2022encodec}. AudioLM shows how discrete audio representations support language-model-style generation~\citep{borsos2023audiolm}. This representation makes separation amenable to conditional sequence prediction, alongside tasks such as synthesis and recognition. TokenSplit is a direct precedent: it jointly handles speech separation, transcription, and speech generation through discrete representations, with both direct separation and refinement of a conventional separator's outputs, and with or without transcript conditioning~\citep{erdogan2023tokensplit}.

\modelname{} uses codec sequences as its separation output, conditioned on continuous audio features and discrete mixture codes (\cref{sec:model}). Its backbone predicts the response structure and conditions a local decoder for each frame's codec codes. The relevant distinction is the combination of variable voice-stream generation, speaker-timeline supervision, and adaptation on real mixtures through candidate feedback. Generating plausible audio and preserving the particular voices and words in a mixture remain separate requirements, motivating the content and identity checks used in our evaluation and application.

\subsection{Prompted Extraction and Singing-Voice Separation}
Target extraction specifies which sound to recover through an external cue. AudioSep uses natural-language descriptions~\citep{liu2023audiosep}; SoloAudio performs generative extraction with a diffusion Transformer and supports audio and language conditioning~\citep{wang2024soloaudio}; and SAM Audio supports text, visual, and temporal prompts~\citep{samaudio2025}. These interfaces allow a user to select a sound of interest. Blind voice separation instead requires enumerating the audible voices without a target query. Our comparisons make the supplied-count setting explicit because count information and target selection change what a system must infer.

Music separation introduces sustained pitch, accompaniment, and potentially layered vocals. Mel-Band RoFormer models overlapping mel-frequency bands for music source separation~\citep{wang2023melband}. MIR-1K and MUSDB18-HQ provide established singing and music evaluation resources~\citep{hsu2010mir1k,rafii2019musdb18hq}. We use a fixed music separator to produce song pseudo-vocals and train these as one vocal stream (\cref{sec:data}). This supervision can represent multiple singers in a combined vocal target, whereas speech examples teach individual voice streams. The two target definitions determine what each evaluation can establish.

\subsection{Learning from Unpaired Recordings and Teacher Targets}
MixIT removes the need for isolated training sources by separating a mixture of mixtures and assigning estimated components back to the observed mixtures~\citep{wisdom2020mixit}. Its application to real meetings demonstrates adaptation to far-field, overlapping, reverberant speech without isolated in-domain references~\citep{sivaraman2021mixit}. RemixIT follows a teacher--student approach: estimated speech and noise are remixed to create training pairs, and the teacher is updated from successive students~\citep{tzinis2022remixit}. These methods provide alternatives to relying entirely on constructed mixtures with known sources.

Our supervision is organized by the reference available for each recording. Constructed mixtures provide source audio, songs provide fixed-teacher pseudo-vocals, and natural recordings provide speaker timelines. Subsequent RL compares candidate separations of the same real mixture using a model judge (\cref{sec:training}). The distinction concerns the training signal: mixture reconstruction, teacher-generated audio, speaker-activity labels, and relative quality feedback impose different constraints on the outputs.

\subsection{Speaker Activity, Diarization, and Identity}
Diarization estimates who speaks when, including overlapping activity. End-to-end neural diarization formulates this as multi-label prediction with a permutation-free objective and supports adaptation from speaker-segment labels~\citep{fujita2019eend}. Such labels describe temporal activity without specifying each speaker's waveform. MOSS-Transcribe-Diarize supplies the speaker-attributed intervals used in our timeline supervision~\citep{moss_transcribe_diarize_2026}. We encode those intervals as a text response within the same model that generates separated audio, allowing natural recordings to contribute supervised examples without isolated targets.

Activity labels and speaker identity serve different roles in long recordings. Speaker-verification systems such as CAM++ and ERes2Net learn representations for comparing voices~\citep{wang2023campp,chen2023eres2net}. Our conversation pipeline uses speaker embeddings and consistency across pieces to associate local outputs with recording-level speakers, then combines activity and identity checks to detect missing or intruding speech (\cref{app:duplex}). Agreement on identity alone does not establish that the original words have been preserved, which motivates the separate content checks.

\subsection{AI Feedback, Policy Optimization, and Judge Reliability}
RLHF uses human preferences to train language models beyond supervised demonstrations~\citep{ouyang2022rlhf}, while RLAIF studies preferences supplied by AI systems~\citep{lee2024rlaif}. PPO supplies a clipped policy-gradient objective~\citep{schulman2017ppo}; GRPO in DeepSeekMath and the optimization components of DAPO develop group-based feedback and practical training procedures for language-model RL~\citep{shao2024deepseekmath,yu2025dapo}. Our update uses relative rewards within a group of candidate separations, with an unnormalized group-centered advantage and signal-based shaping (\cref{sec:rl}).

Audio-specific feedback learning provides closer precedents. MOS-based RLHF fine-tunes speech enhancers using a learned perceptual-quality reward~\citep{kumar2025speechrlhf}. MARS-Sep optimizes a mask policy with a multimodal preference reward for query-conditioned sound separation~\citep{zhang2026marssep}. LLM-guided audio-visual enhancement converts audio-language descriptions into scalar ratings for PPO updates~\citep{chen2026llmguided}. These studies connect feedback to audio processing through quality prediction, cross-modal alignment, or textual descriptions. Our setting combines blind generation of multiple voice streams with comparison of complete candidate separations of the same recording. The feedback must account for voice purity and coverage across all outputs, as well as each track's quality.

The quality of the feedback is a separate concern from the optimizer. LLM-as-a-judge studies identify presentation-order bias and other limitations in model judgments~\citep{zheng2023llmjudge}. Reward-model overoptimization can increase a proxy score while degrading the quality it is intended to measure~\citep{gao2023overoptimization}. Our judge first describes clips independently, then compares their reports alongside an anonymous mixture anchor and filters inconsistent rankings under shuffled and reversed candidate orders. The consistency check targets sensitivity to candidate order, and the signal checks target degenerate outputs.

\subsection{Evaluating Content, Acoustics, Pitch, and Identity}
Waveform-based measures such as BSS Eval and SI-SDR quantify aspects of reconstruction fidelity relative to source signals~\citep{vincent2006bsseval,leroux2019sisdr}. Recognition-based evaluation instead asks whether the separated speech retains recoverable words. Permutation-aware multi-speaker ASR and work on WER definitions provide the basis for matching streams before counting transcription errors~\citep{qian2017multitalker,sklyar2021streaming,vonneumann2023wer}. Our PI-WER protocol pads unequal stream sets with empty transcripts and evaluates the minimum-cost assignment; count accuracy separately assesses the number of voiced outputs.

Speech representations such as WavLM are trained to support tasks involving both content and speaker information~\citep{chen2022wavlm}. Our SingerSim metric uses WavLM features, while raw pitch accuracy measures pitch preservation (\cref{app:evaluation:music}). These quantities complement transcription recovery: correct words do not imply correct pitch or voice identity, and high embedding similarity need not imply exact words. Reference provenance also matters. Comparisons against teacher-extracted vocals measure agreement with that teacher, while isolated-source references support a different fidelity interpretation. We therefore distinguish these dimensions and reference types rather than treating one score as a complete measure of separation.

\subsection{Continuous Separation and Full-Duplex Conversation Data}
Continuous speech separation converts long recordings containing overlap into continuous output channels suitable for recognition. Speaker-independent CSS allows successive utterances to occupy different channels while keeping each utterance intact~\citep{yoshioka2019css}. Our application requires an additional association step: independently generated pieces must be linked to persistent speaker identities and placed on the original recording timeline. Window planning, stream assignment, alignment, and targeted repair jointly determine whether the assembled tracks preserve coverage (\cref{sec:duplex}).

Full-duplex dialogue models such as Moshi model user and system speech in parallel streams to support overlap and conversational timing~\citep{defossez2024moshi}. This creates a need for temporally aligned conversational audio. Our pipeline prepares continuous speaker tracks and aligned transcripts offline from existing recordings. It checks speaker purity, missing speech, and content consistency before packaging the outputs.

\FloatBarrier
\section{Generative Blind Voice Separation}\label{sec:task}
We formulate blind voice separation as conditional audio generation: given a single-channel recording, the model infers how many voices are present and generates a separate waveform for each. The recording may also contain music or environmental sound, and inference requires neither a supplied voice count nor a target-speaker query. \modelname{} implements this task by autoregressively generating a variable number of voice streams as sequences of audio-codec frames, which are decoded into waveforms. Its output contains only the separated voices, with no background or residual stream.

\subsection{Task and Representation}\label{app:model}
For a single-channel recording $x$ containing $K\geq1$ voices, a response contains a declared count $\Khat$ and $J$ generated streams:
\begin{equation}
 (\Khat,\hat y_{1:J})\sim\pi_\theta(\cdot\mid x),
 \qquad \Khat\in\{1,\ldots,\Kmax\},
 \label{eq:task}
\end{equation}
Here $\Kmax$ is the decoding cap. The declared count and emitted stream count are distinct so that malformed responses can be measured. The vocabulary covers voice-containing recordings and has no zero-voice response. With $T$ input frames and $T_k$ frames in output stream $k$, each codec frame contains twelve codes $a_{tq}\in\{0,\ldots,1023\}$; code 1024 denotes padding. The signal gate additionally computes a valid-stream count $J_{\mathrm{eff}}$ (\cref{app:judge:constants}). Signal validity is independent of whether a stream isolates the correct speaker.

\begin{figure}[htbp]
\centering
\includegraphics[width=\linewidth]{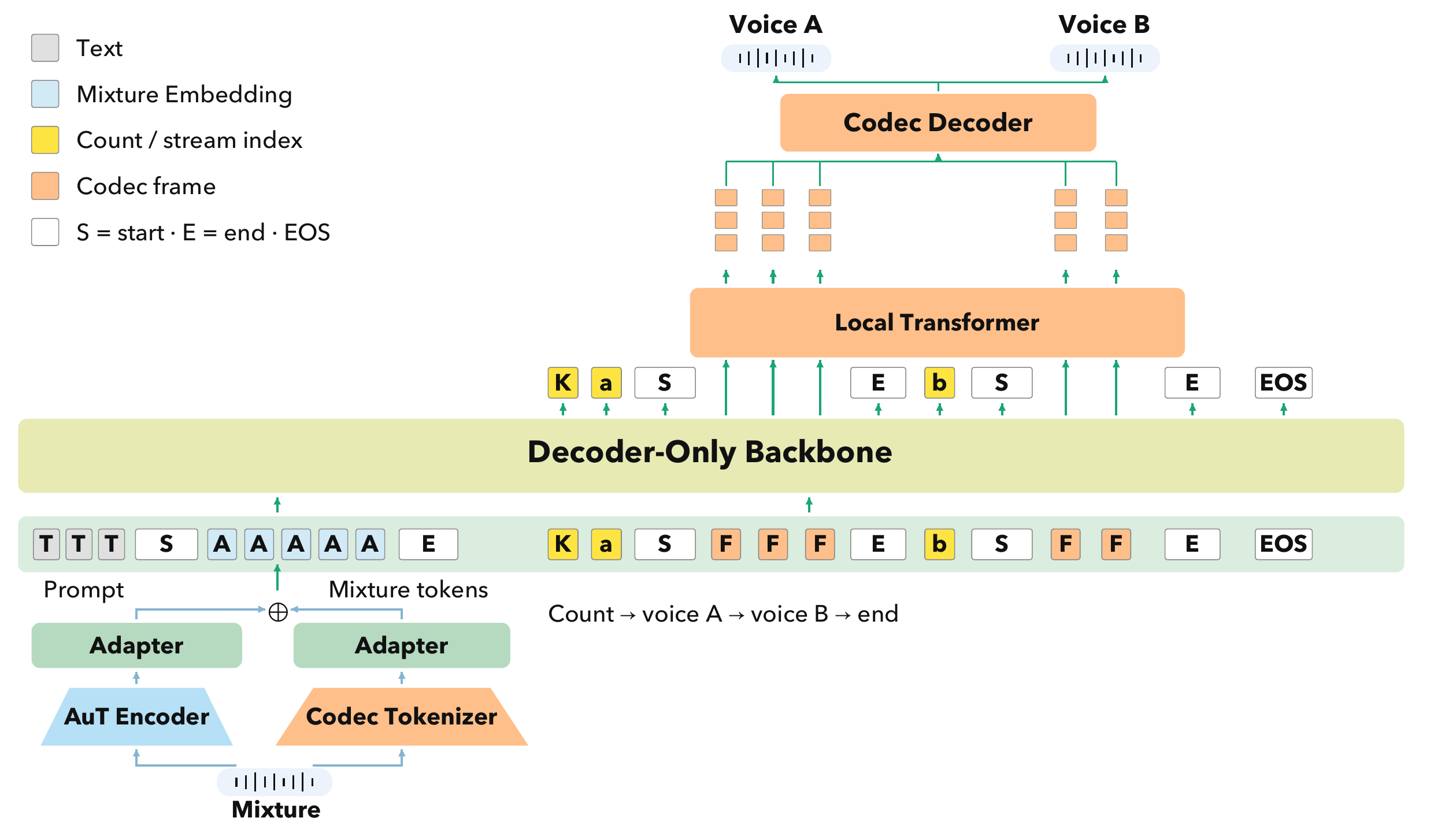}
\caption{Generative separation architecture of \modelname{}. Continuous audio features and discrete mixture codes condition the backbone. The text head predicts control tokens, and a local transformer generates the codes within each audio frame.}
\label{fig:system}
\end{figure}

\subsection{Architecture}\label{sec:model}
\modelname{} uses a Qwen3-4B decoder~\citep{yang2025qwen3} conditioned on both continuous and discrete representations of the mixture (\cref{fig:system}). The AuT encoder from Qwen3-Omni~\citep{qwen3omni} supplies continuous features, and MOSS-Audio-Tokenizer-v2~\citep{gong2026mosstokenizer} supplies twelve residual codec codes per frame at 12.5 frames/s. Gated multilayer perceptrons project the two representations into the backbone space and sum them at input frame positions. At an output frame, a one-layer local transformer predicts the twelve codes sequentially, conditioned on the backbone state. This frame-local decoder contains causal multi-head self-attention and a two-layer feed-forward network, each preceded by layer normalization and wrapped in a residual connection. Codebook-specific linear heads predict each code using the backbone state and the preceding codes within that frame. A separate adaptor feeds the generated codes back into the backbone. The codec is fixed; the audio encoder, backbone, adaptors, and local transformer are trained.

The full component configuration is listed in \cref{tab:config}.

\subsection{Output Format and Inference}
The backbone's text head predicts tokens that control the output structure, including the voice count, stream indices, and audio boundaries. For separation, it first emits the count and the first stream index, followed by an audio-start token to begin that stream. Whenever the text head emits a frame-request token, the corresponding backbone hidden state conditions the local transformer, which generates the twelve codec codes for that audio frame. The output adaptor maps these codes back into the backbone input space for subsequent autoregressive steps. Frame generation repeats until the text head emits an audio-end token. The model then generates the next stream index and its audio, or emits an end-of-response token to end the response. The codec decoder converts each stream's frame sequence into a waveform.

During supervised training, voice targets are ordered by first onset to define a canonical serial output sequence; evaluation still matches generated streams to references by permutation. For the speaker-timeline task, a separate instruction asks the text head to emit $[\mathrm{start}][\mathrm{speaker}][\mathrm{end}]$ triples sorted by start time, without triggering audio-frame generation. Complete prompts and token roles appear in \cref{app:model:prompts}.

\FloatBarrier
\section{Supervised Learning from Multiple Sources}\label{sec:data}
\begin{figure}[htbp]
\centering
\includegraphics[width=\linewidth]{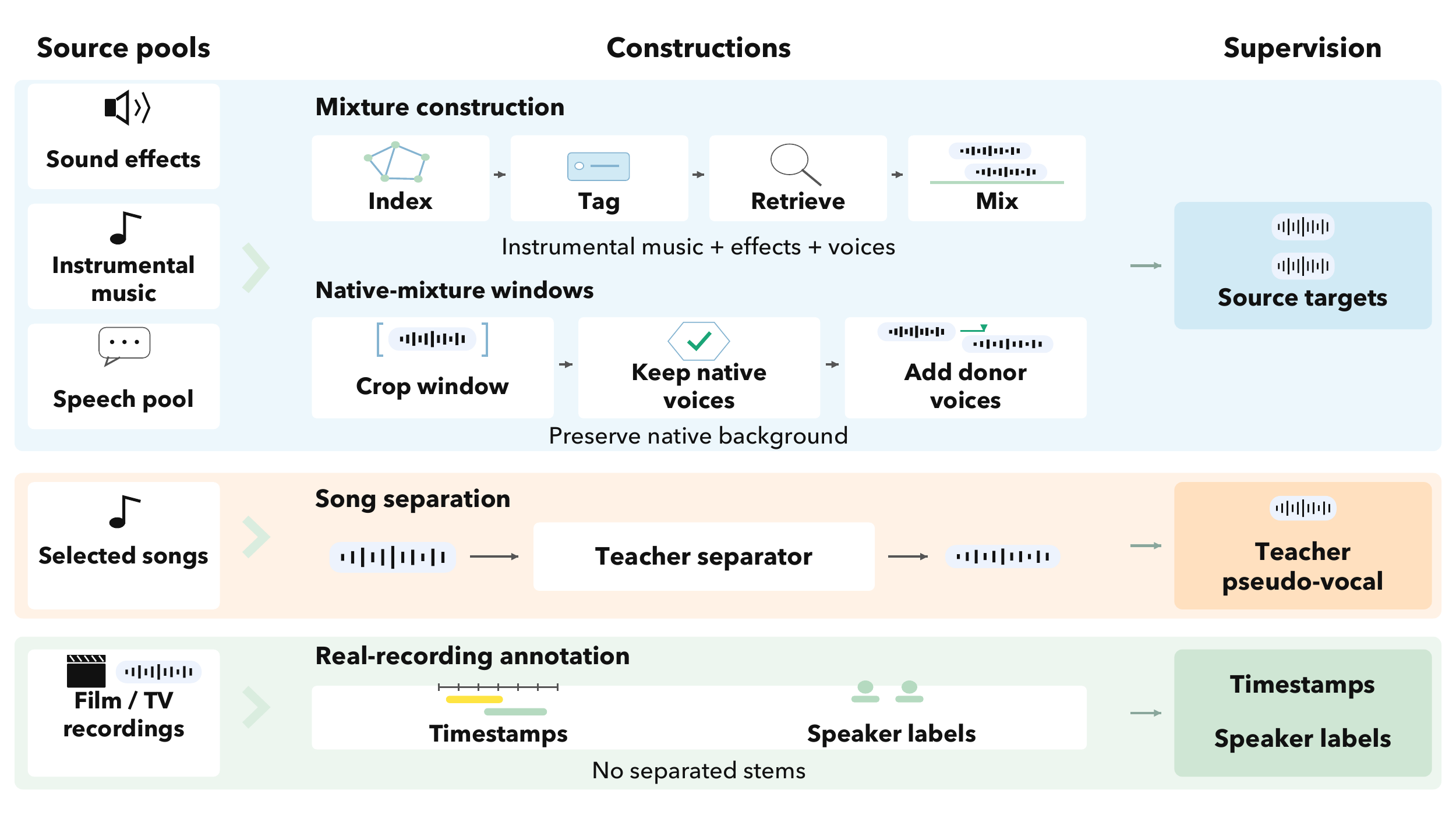}
\caption{Data construction and the resulting supervision. Known-source mixtures supply audio targets, songs supply teacher pseudo-vocals, and annotated recordings supply speaker timelines. The timeline response contains timestamps and speaker labels.}
\label{fig:data}
\end{figure}

The first training stage teaches the model to take a mixed recording as input and serially generate the separated voice streams. To establish this ability, we train a single model with three forms of supervision: source waveforms used to construct speech mixtures, teacher-generated vocal targets for songs, and speaker timelines for natural recordings (\cref{fig:data,tab:data}). The first two supervise audio generation, while the third teaches the model who speaks when through text outputs containing timestamps and speaker labels.

\begin{table}[!htbp]
\caption{Training supervision. Timelines supervise text only; pseudo-vocals inherit teacher limitations. Data construction is detailed in \cref{app:data}.}
\label{tab:data}
\centering\small
\begin{tabularx}{\linewidth}{@{}lYY@{}}
\toprule
Input family & Available target & Supervised response \\
\midrule
Constructed speech mixtures & Source audio from the construction & Count and per-voice codec sequences \\
Songs & Teacher-generated vocal audio & One vocal codec sequence with count and delimiters \\
Natural recordings & Speaker-attributed time intervals & Text-only speaker timeline \\
\bottomrule
\end{tabularx}
\end{table}

\subsection{Data Construction}\label{app:data}
\paragraph{Constructed mixtures.}
We construct mixtures from voice sources, a cleaned Pixabay sound-effect library, and music recordings. Mixtures are built in three steps. \emph{Semantic tagging} labels each background recording with a scene category derived from pre-computed scene topics. \emph{Candidate retrieval} queries each scene label for semantically similar audio and keeps a pool of 20 candidate clips that mixes long and short recordings at a fixed ratio. \emph{Mixture synthesis} lays a long recording as the acoustic bed when a bed is used, overlays short sound effects at random positions, and adds the voice sources as independent streams. Because every source is known by construction, the isolated voice waveforms serve as separation targets without manual annotation. Scene mixtures place one to four Chinese, English, or Japanese voices on a retrieved background: a music or ambient bed, a bed with transient effects, transient effects alone, or silence. Overlay and background-preserving windows instead add donor voices to 5--30 s windows that already carry an acoustic background. Overlay windows contain three to five voices, while background-preserving windows contain two to six.

\paragraph{Speaker timelines.}
Speaker timelines describe who speaks when without providing either words or separated waveforms. These rows train the model to infer speaker activity from natural acoustics through the timeline instruction. They do not provide waveform supervision for separation. We annotate the natural recordings with MOSS-Transcribe-Diarize~\citep{moss_transcribe_diarize_2026}, retaining speaker labels and start/end times to construct the timeline targets.

We select multi-speaker recordings from in-domain data spanning film, television, animation, short video, podcasts, radio, audiobooks, and sports commentary to provide speaker-activity supervision. The nominal windows span 10--120 s and contain two to six speakers. Targets are time-interval triples ordered by start time, with speaker IDs assigned by first appearance. Upstream transcripts are excluded from the prediction target; the output format is specified in \cref{app:model:prompts}.

\paragraph{Singing vocals.}
A fixed Mel-Band RoFormer separator~\citep{wang2023melband} extracts a vocal pseudo-target from each song. Training uses the song as input and the teacher's vocal output as a single target stream. Accompaniment is not an output target. This is teacher supervision, so agreement with the pseudo-target measures consistency with the teacher rather than recovery of an independently recorded studio vocal.

Song segments are cut from in-house song recordings into clips of at most two minutes before teacher extraction. The teacher's vocal stream can contain backing vocals or multiple singers, so it forms a combined vocal target rather than individual singer streams.

\subsection{Supervised Sampling}
The sampler groups examples by source count and duration, with duration boundaries at 10 and 20 s. Timelines form a separate administrative class indexed by zero; their shortest-duration cell has zero weight, leaving twenty active classes. The sampling-budget shares are 4.81\% for timelines and 4.0/26.6/39.5/19.8/4.5/0.8\% for one through six source streams.

\subsection{Supervised Objective and Optimization}\label{sec:sft}\label{app:training}\label{app:training:lineage}\label{app:training:sft}
Let $z$ contain the assistant text-channel tokens, including frame requests and delimiters, and let $a_{tq}$ denote codebook $q$ of output frame $t$. The supervised objective combines text and audio prediction:
\begin{equation}
 \mathcal L_{\mathrm{sup}}=\mathcal L_{\mathrm{text}}+\lambda\mathcal L_{\mathrm{codec}},\qquad \lambda=1.
 \label{eq:sft}
\end{equation}
Both terms are teacher-forced cross-entropies. The codec term is absent for timeline rows, while the text term supervises the count, stream boundaries, and frame requests for audio-producing rows. No additional count or boundary loss is used.

Text positions $u$ and output codebook positions $(t,q)$ contribute the negative log probabilities
\begin{equation}
 -\log p_\theta(z_u\mid z_{<u},x),\qquad
 -\log p_\theta(a_{tq}\mid a_{t,<q},h_t),
\end{equation}
where $h_t$ conditions output frame $t$.

Supervised training uses a peak learning rate of $10^{-4}$. It establishes separation and speaker-activity prediction before the real-recording RL stage in \cref{sec:training}.

\FloatBarrier
\section{Reinforcement Learning on Real Recordings}\label{sec:training}
Supervised training establishes the separation interface using source audio and speaker timelines. Reinforcement learning then adapts the model to real recordings without isolated source targets. For each recording, we sample alternative separations and use a model judge to compare voice purity, content coverage, and audio quality.

\subsection{RL Data}\label{app:data:rl}
The RL prompt set consists of real recordings without isolated source targets. Every recording has a voice-count annotation. AliMeeting~\citep{yu2022m2met} uses its manual TextGrid annotations; other recordings use dataset annotations where available and MOSS-Transcribe-Diarize otherwise. All prompts use the model-judge reward described in \cref{sec:judge}, with the annotated count used in the signal checks and ranking request.

The recordings include far-field meetings, film, television, animation, podcasts, short video, and audiobooks.

\subsection{Model-Judge Reward on Real Recordings}\label{sec:judge}
\begin{figure}[htbp]
\centering
\includegraphics[width=\linewidth]{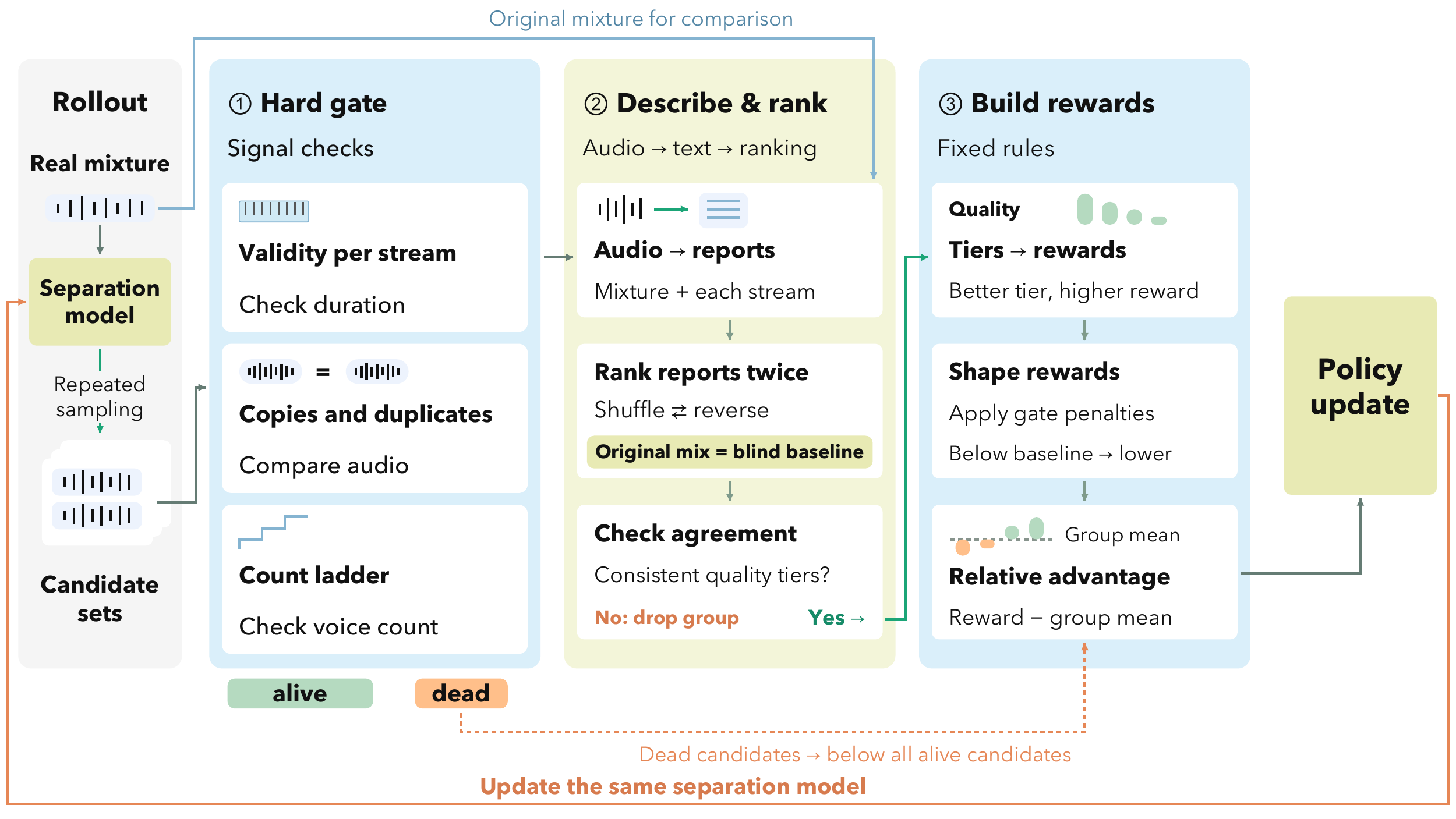}
\caption{Judge feedback for a group of candidate separations. Signal checks screen stream validity, copies, duplicates, and count errors. An audio-language model describes clips individually, and a text ranker compares their reports with an anonymous mixture anchor in shuffled and reversed orders. Groups that fail the order-consistency check are dropped. Accepted groups receive shaped rewards, with dead candidates below surviving ones; subtracting the group mean gives the policy advantage.}
\label{fig:judge}
\end{figure}

In early trials, directly asking Qwen3-Omni to assign rubric scores to separated audio produced large score fluctuations across repeated evaluations of the same clip. We therefore construct relative feedback in two steps (\cref{fig:judge}): an audio-language model first describes the mixture and each candidate track independently, and a text model then ranks candidate sets from these reports, using an anonymous copy of the input mixture as an anchor. Repeating the ranking under reversed candidate order provides a consistency check for selecting feedback used in policy updates.

\paragraph{Signal checks.}
We screen stream duration, signal level, and active fraction, then detect likely input copies and duplicate streams with a log-mel frame-match score. Let $J_{\mathrm{eff},i}$ count the streams in candidate $i$ that pass the duration, level, and activity checks. Given the count annotation $K$, $J_{\mathrm{eff},i}-K$ determines whether to penalize excess or missing streams. A candidate is marked dead when it emits no usable stream, exceeds the duration bound, receives a zero count penalty, or only passes through the mixture under the applicable copy rule. These checks are heuristics; perceptually incorrect outputs can still pass them.

\paragraph{Independent perception and an input anchor.}
Qwen3-Omni-30B-A3B-Instruct~\citep{qwen3omni} describes the mixture and each surviving track in separate calls. Reports contain audible speaker count, voice traits, content, artifacts, and silence. All clips in a group use the same temporal window. The input mixture is also described with the track prompt and inserted anonymously as a candidate. This anchor lets the ranker compare a separation attempt with returning the input unchanged.

\paragraph{Repeated ranking and consistency.}
Qwen3.6-27B~\citep{qwen36} receives text only: the mixture report, the per-track reports of every candidate, and the count annotation $K$, without audio input. It assigns candidates to four quality tiers, prioritizing single-voice purity, coverage of the mixture's speakers and content, avoidance of duplicates and invented content, and audio quality. Both ranking calls use the same descriptions and anonymous mixture anchor: candidate labels are shuffled in the first call and presented in reverse order in the second. After mapping results back to the original candidates, let $v_i^{(m)}$ be candidate $i$'s tier score in call $m$. We compute
\begin{equation}
 v_i^{(m)}\in\{q_A,q_B,q_C,q_D\},\qquad
 t_i=\frac{1}{M}\sum_{m=1}^{M}v_i^{(m)},
 \label{eq:rankmean}
\end{equation}
where $M$ is the number of ranking calls and $q_A>q_B>q_C>q_D$ are the tier values. The anonymous anchor is scored in the same way. Their settings and the agreement threshold $\tau_{\min}$ are given in \cref{tab:judgeconfig}.

Since presentation order does not change audio quality, large preference changes after reversal indicate order-dependent or unstable feedback. We measure agreement between the two rankings $\mathcal R^{(1)}$ and $\mathcal R^{(2)}$ with Kendall's tie-aware correlation $\tau_b$~\citep{kendall1945ties}:
\begin{equation}
 \tau_b\!\left(\mathcal R^{(1)},\mathcal R^{(2)}\right)<\tau_{\min}
 \quad\Longrightarrow\quad r_i=0\quad\forall i\text{ in the group}.
 \label{eq:rankfilter}
\end{equation}
This rule permits small ranking changes but discards the entire group's feedback when agreement falls below the threshold. We also reject groups when both calls cannot distinguish candidates or disagree on whether they are distinguishable. Rejected groups have zero reward spread and are excluded by the subsequent dynamic filter, so they do not contribute to RL updates. Groups that pass the check use the mean of the two tier scores. The consistency check tests sensitivity to candidate order; the accuracy of content judgments remains dependent on the audio descriptions. Inference settings and failure handling are specified below; \cref{app:judge:validation} tests the judge on known-source mixtures, and \cref{app:judge} provides the signal constants and exact prompts.

\Needspace{8\baselineskip}
\paragraph{Reward shaping.}
Let $t_i$ be candidate $i$'s mean tier score, $t_a$ the anchor score, and $\gamma_i\in[0,1]$ its signal-check multiplier. The multiplier combines the mean per-stream copy/duplicate cap with duration, count, and declared-versus-emitted count penalties. We assign
\begin{equation}
 \Rjudge(i)=
 \begin{cases}
 \delta\left(\eta+(1-\eta)s_i/s_{\max}\right), & i\text{ is dead},\\
 \max(\epsilon_{\mathrm{alive}},\min(1,t_i\gamma_i)), & i\text{ survives},\ t_i\geq t_a,\\
 \max(f_{\mathrm{below}},\beta_{\mathrm{below}}t_i\gamma_i), & i\text{ survives},\ t_i<t_a.
 \end{cases}
 \label{eq:judge}
\end{equation}
Here $\delta$ bounds dead-candidate rewards, $\eta$ sets their minimum fraction, and $s_i$ takes a discrete value determined by the failure category, with maximum $s_{\max}$. The surviving-candidate guard $\epsilon_{\mathrm{alive}}$ ensures that every surviving reward exceeds the dead-candidate maximum. The below-anchor branch has its own floor $f_{\mathrm{below}}$ and scale $\beta_{\mathrm{below}}$. These floors serve different purposes and need not preserve the ranker's ordering; signal penalties can also change that ordering. Parameter values, failure categories, and multiplier details appear in \cref{app:judge:constants}.

\subsection{Judge Inference and Failure Handling}\label{sec:judge:inference}
Each perception call processes one clip. For long recordings, the mixture, anchor, and candidate streams share a uniformly positioned 30 s window. Mixture descriptions provide global context; independently obtained track descriptions provide the evidence for each candidate. The mixture anchor uses the same track prompt as generated streams. The text ranker receives the mixture report, the candidate track reports, and the count annotation $K$; it receives no audio. The perception requests do not receive the count annotation.

\FloatBarrier

The serving backend disables thinking through the chat-template option. If the primary text service fails, the implementation falls back to the perception model for text ranking. This branch uses a different ranker. A description fails when the perception model returns no valid JSON in three attempts. A surviving candidate with a failed track description is left out of the ranking and receives a floor reward of 0.05; if fewer than two fully described candidates remain, the whole group is rejected.

\subsection{Per-Stream Feedback}\label{app:judge:stream}
The track report also supplies local feedback for audio-token updates. For stream $k$ in candidate $i$, let $n_{ik}$ denote the audible-speaker count, $\alpha_{ik}\in\{0,0.05,0.2\}$ the penalty for none, mild, or severe artifacts, and $\sigma_{ik}$ the mostly-silence indicator. The stream score is
\begin{equation}
 q_{ik}=\clip\!\left(1-0.35\min(\max(n_{ik}-1,0),2)-\alpha_{ik}-0.3\sigma_{ik},0,1\right)\mathrm{cap}_{ik}.
 \label{eq:stream}
\end{equation}
The score uses $n_{ik}=1$ for a missing count field and defaults to 0.5 before the cap when the entire report is missing. The trainer centers stream scores and adds them to the corresponding audio-token advantages with coefficient 0.5.

\subsection{Policy Optimization}\label{sec:rl}
We sample $G$ candidates per real-recording prompt and center the judge rewards $r_i=\Rjudge(i)$ within each group:
\begin{equation}
 A_i=r_i-\frac{1}{G}\sum_{j=1}^{G}r_j.
 \label{eq:advantage}
\end{equation}
The update uses a clipped policy-gradient objective~\citep{schulman2017ppo,shao2024deepseekmath}, with token averaging, asymmetric clipping, and dynamic group filtering related to recent language-model RL methods~\citep{yu2025dapo}. We do not divide advantages by the group standard deviation. Reward magnitudes therefore affect the update, as well as candidate order. Groups with reward standard deviation at most $\sigma_{\min}$ are filtered to exclude weakly differentiated groups; small nonzero spread would otherwise yield nonzero advantages. A centered per-stream signal, weighted by $\lambda_{\mathrm{stream}}$, further shapes audio-token advantages. The training loss combines the resulting clipped policy loss with KL regularization and an auxiliary codec cross-entropy term:
\begin{equation}
 \mathcal L=\mathcal L_{\mathrm{PG}}+\beta_{\mathrm{KL}}\mathcal L_{\mathrm{KL}}+\lambda_{\mathrm{CE}}\mathcal L_{\mathrm{CE}}^{+}.
 \label{eq:rl-loss}
\end{equation}
RL has two phases: 130 updates at a learning rate of $10^{-6}$ without the KL term ($\beta_{\mathrm{KL}}=0$), followed by 90 updates at $2\times10^{-6}$ with it. Training runs in segments, and the KL reference is the policy at the start of the current segment. The codec term applies to positive-advantage rollouts. The clipped surrogate and hyperparameters are specified in \cref{app:training:rl}.

\FloatBarrier
\section{Evaluation}\label{sec:eval}

\subsection{Evaluation Setup}\label{app:evaluation}
We evaluate separation in three domains: overlapping speech, noisy environments, and singing. Baselines comprise the speech separators SepFormer~\citep{subakan2021sepformer}, MossFormer~\citep{zhao2023mossformer}, ConvTasNet~\citep{luo2019convtasnet}, and SR-CorrNet~\citep{shin2026srcorrnet}, and the query-based extractors SAM Audio~\citep{samaudio2025}, AudioSep~\citep{liu2023audiosep}, and SoloAudio~\citep{wang2024soloaudio}, whose target and residual outputs are scored as two streams. With supplied counts, we prefill \modelname{}'s first output token with \texttt{2} for LibriMix, or \texttt{1} for Voices in the Wild (ViW) and singing vocals, matching baselines with fixed output counts or a single target.

\subsubsection{Evaluation Data}\label{app:evaluation:provenance}
\paragraph{LibriMix.}
Clean and noised two-speaker mixtures are generated with the official LibriMix scripts~\citep{cosentino2020librimix}.

\paragraph{Voices in the Wild (ViW) subset.}\label{app:evaluation:viw}
We select 3,000 single-speaker clips from Voices-in-the-Wild-2M~\citep{xie2026megaasr}. From its 54 compound acoustic scenarios, we retain twelve: five single conditions (\texttt{noise}, \texttt{echo}, \texttt{far\_field}, \texttt{obstructed}, and \texttt{recording}) and seven pairs (\texttt{echo\_noise}, \texttt{echo\_recording}, \texttt{far\_field\_noise}, \texttt{far\_field\_recording}, \texttt{obstructed\_noise}, \texttt{obstructed\_recording}, and \texttt{recording\_noise}). Scenarios involving \texttt{distortion} or \texttt{dropout} are excluded because they corrupt or remove speech content beyond the denoising and over-splitting behavior evaluated here. The retained conditions cover noise, reverberation, distance, occlusion, and recording-device artifacts.

We require decodable audio, a duration of 1--30~s, and a nonempty transcript, then randomly sample 250 clips per scenario (125 Mandarin and 125 English) with a fixed seed. Audio is resampled to 16~kHz mono. The resulting set spans 1.0--19.65~s per clip (median 5.11~s), totaling 5.54~h.

\paragraph{WildTV.}
WildTV consists of 46 real-world speech clips from several domains, including film and TV drama and podcasts, each with one to three speakers: 11 with one speaker, 25 with two, and 10 with three. Reference labels (speaker counts and per-speaker transcripts) are annotated with MOSS-Transcribe-Diarize~\citep{moss_transcribe_diarize_2026} and then manually reviewed and corrected. \modelname{} and SR-CorrNet (the released SR-CorrNet-SS checkpoint) each use a single model with blind count inference. SepFormer, MossFormer, and ConvTasNet switch checkpoints using the true count: enhancement for one speaker, Libri2Mix for two, and WSJ0Mix for three. Their perfect output-count agreement follows from this construction.

\paragraph{Singing benchmarks.}
SongYT contains 3,000 song recordings in Chinese, English, and Japanese, so it tests whether separation preserves vocals across languages and regional singing styles. SongYT uses vocal pseudo-references produced by Mel-Band RoFormer~\citep{wang2023melband}. Agreement with these references measures consistency with the teacher estimate. MIR-1K~\citep{hsu2010mir1k} contains 1,000 clips of 4--13 s cut from 110 Chinese karaoke pop songs sung by 19 mostly amateur singers. Accompaniment and singing voice are recorded on the left and right channels, so every clip has an isolated vocal reference. MUSDB18-HQ~\citep{rafii2019musdb18hq} is the uncompressed version of MUSDB18: 150 full-length stereo tracks of various genres (about 10 h), each with isolated vocals, drums, bass, and other stems. MUSDB18-HQ has no lyric references, so we report only RPA and SingerSim on it.

\subsection{Metrics and Aggregation}\label{app:evaluation:metrics}
To measure how well spoken content is preserved, we transcribe the separated audio and compute permutation-invariant WER (PI-WER)~\citep{qian2017multitalker,sklyar2021streaming}. Given $K$ reference transcripts and $J$ output transcripts, we pad the shorter list to $L=\max(K,J)$ with empty transcripts to obtain $\tilde r$ and $\tilde h$, then compute
\begin{equation}
\mathrm{PI\text{-}WER}
=\min_{\pi\in\mathcal{P}_L}
\frac{\sum_{k=1}^{L} d_{\mathrm{edit}}(\tilde r_k,\tilde h_{\pi(k)})}
     {\sum_{k=1}^{K}|r_k|}.
\label{eq:piwer}
\end{equation}
Here, $\mathcal{P}_L$ contains all stream permutations, $d_{\mathrm{edit}}$ is token edit distance, and the denominator counts all reference tokens. We find the minimum-cost one-to-one assignment with the Hungarian algorithm~\citep{vonneumann2023wer}.

When counts match, PI-WER removes arbitrary output ordering; with one reference and one output, it reduces to ordinary WER. ViW and singing have one reference, so we report WER; for baselines that return two streams, the stream with the lower WER is matched to the reference. For multi-speaker and blind evaluation, all emitted streams enter the padded assignment: when counts differ, unmatched reference tokens count as deletions and unmatched output tokens as insertions. Extra output can therefore push WER above 1, while an extra empty transcript adds no errors.

\paragraph{Transcription and PI-WER.}
Separated outputs are resampled to 16 kHz and transcribed with Whisper large-v3~\citep{radford2023whisper}. Text is normalized with Unicode NFKC and case folding. CJK characters, kana, and Hangul are tokenized character by character, while Latin text is tokenized into words. Punctuation separates tokens, except that internal apostrophes are retained. Thus, the reported WER uses script-dependent word/character units. Count accuracy compares the number of voiced output streams with the reference speaker count; the declared count is stored separately as \texttt{stated}.

Corpus-level WER is a micro-average: total edit errors divided by total reference tokens, rather than the mean of clip-level WERs.

\paragraph{Speech quality.}
We use DNSMOS on ViW to measure how much background noise remains after separation. ViW clips carry noise, reverberation, and other acoustic degradations, and WER alone cannot show whether a system removes the noise or only keeps the speech intelligible. DNSMOS~\citep{reddy2022dnsmos} is a non-intrusive neural predictor of ITU-T P.835 subjective ratings. Its background-noise score (BAK) rates how intrusive the remaining background is, and its overall score (OVRL) rates overall quality; both use a 1--5 scale with higher values better. We report means and medians over the ViW clips.

\paragraph{Music similarity and pitch.}\label{app:evaluation:music}
SingerSim uses layer-14 features from WavLM-large~\citep{chen2022wavlm}, with L2 normalization on RMS-active frames followed by mean pooling. RPA (raw pitch accuracy) is the fraction of voiced reference frames whose estimated F0 lies within 50 cents of the reference F0. Transcript-based metrics are insensitive to melody: a separated vocal can keep every lyric while its pitch drifts. This matters for a generative separator, which re-synthesizes the vocal rather than masking the mixture. RPA therefore tests whether separation preserves the sung melody, complementing WER for lyric content and SingerSim for vocal similarity.

\subsection{Overlapping Speech} LibriMix~\citep{cosentino2020librimix} mixes two speakers, with and without added noise (\cref{tab:librimix}). WildTV contains 46 natural recordings with one to three speakers (\cref{tab:wildtv}). On WildTV, SepFormer, MossFormer, and ConvTasNet use a separate checkpoint for each speaker count, selected by the true count; \modelname{} and SR-CorrNet each infer the count with a single model.

\begin{table}[htbp]
\centering
\caption{LibriMix PI-WER $\downarrow$ on overlapping speech with the count supplied. Bold/underline: best/second.}
\label{tab:librimix}
\small
\setlength{\tabcolsep}{7pt}
\begin{tabular}{@{}lrr@{}}
\toprule
System & Clean & Noised \\
\midrule
SepFormer & \textbf{0.058} & \underline{0.218} \\
MossFormer & 0.080 & 0.289 \\
SAM Audio & 0.156 & 0.351 \\
SR-CorrNet & 0.165 & 0.682 \\
AudioSep & 0.435 & 0.568 \\
ConvTasNet & 0.493 & 0.824 \\
SoloAudio & 0.615 & 0.854 \\
\midrule
\modelname{} & \underline{0.060} & \textbf{0.192} \\
\bottomrule
\end{tabular}

\end{table}
\begin{table}[htbp]
\centering
\caption{WildTV corpus-level PI-WER $\downarrow$ and blind count accuracy (Acc.) $\uparrow$. Dashes: inapplicable. Bold/underline: best/second.}
\label{tab:wildtv}
\small
\setlength{\tabcolsep}{7pt}
\begin{tabular}{@{}lrrrrr@{}}
\toprule
& \multicolumn{4}{c}{PI-WER} & \\
\cmidrule(lr){2-5}
System & Overall & 1 spk & 2 spk & 3 spk & Acc. \\
\midrule
\multicolumn{6}{l}{\textit{Non-blind}} \\
MossFormer & \underline{0.527} & \underline{0.032} & \underline{0.371} & \underline{1.386} & --- \\
SepFormer & 0.967 & 0.783 & 0.801 & 1.575 & --- \\
ConvTasNet & 1.541 & 0.468 & 0.939 & 4.113 & --- \\
\midrule
\multicolumn{6}{l}{\textit{Blind}} \\
SR-CorrNet & 1.940 & 1.572 & 1.864 & 2.471 & \underline{0.217} \\
\modelname{} & \textbf{0.150} & \textbf{0.025} & \textbf{0.160} & \textbf{0.238} & \textbf{1.000} \\
\bottomrule
\end{tabular}

\end{table}
\FloatBarrier
\subsection{Speech in Noisy Environments} On ViW~\citep{xie2026megaasr} (\cref{tab:viw}), separation should remove noise and other interference while keeping the voice intact. The unprocessed input yields 0.261 WER; every separator raises it, \modelname{} the least (0.302), while \modelname{} obtains the highest DNSMOS background (BAK) and overall (OVRL) quality.

\begin{table}[htbp]
\centering
\caption{ViW WER $\downarrow$ and DNSMOS~\citep{reddy2022dnsmos} BAK/OVRL $\uparrow$ (mean/median). Bold/underline: best/second separator.}
\label{tab:viw}
\small
\setlength{\tabcolsep}{7pt}
\begin{tabular}{@{}lrrr@{}}
\toprule
System & WER & BAK & OVRL \\
\midrule
Unprocessed input & 0.261 & 1.63/1.25 & 1.43/1.16 \\
\midrule
SepFormer & 0.847 & 1.87/1.31 & 1.54/1.19 \\
MossFormer & \underline{0.315} & \underline{2.28}/1.71 & 1.69/1.23 \\
SAM Audio & 0.708 & 2.26/\underline{2.06} & \underline{1.77}/\underline{1.60} \\
SR-CorrNet & 1.074 & 1.67/1.59 & 1.56/1.55 \\
AudioSep & 0.540 & 1.94/1.53 & 1.60/1.35 \\
ConvTasNet & 1.082 & 1.56/1.53 & 1.53/1.52 \\
SoloAudio & 0.882 & 1.93/1.38 & 1.52/1.24 \\
\midrule
\modelname{} & \textbf{0.302} & \textbf{3.18/3.45} & \textbf{2.45/2.56} \\
\bottomrule
\end{tabular}

\end{table}
\FloatBarrier
\subsection{Singing Vocals} We evaluate singing vocals on SongYT, MIR-1K~\citep{hsu2010mir1k}, and MUSDB18-HQ~\citep{rafii2019musdb18hq}, additionally reporting raw pitch accuracy (RPA) and vocal similarity (SingerSim) (\cref{tab:music}).

\begin{table}[!htbp]
\centering
\caption{Singing results: WER $\downarrow$, mean RPA $\uparrow$, and mean SingerSim $\uparrow$. Bold and underline mark the best and second-best comparable results. MUSDB18-HQ has no lyric references. Unprocessed mixtures are omitted: Whisper large-v3 gives very high WER on the original music, so the comparison is not informative.}
\label{tab:music}
\small
\setlength{\tabcolsep}{3pt}
\begin{tabular*}{\linewidth}{@{\extracolsep{\fill}}lrrrrrrrr@{}}
\toprule
& \multicolumn{3}{c}{SongYT} & \multicolumn{3}{c}{MIR-1K} & \multicolumn{2}{c}{MUSDB18-HQ} \\
\cmidrule(lr){2-4}\cmidrule(lr){5-7}\cmidrule(lr){8-9}
System & WER & RPA & SingerSim & WER & RPA & SingerSim & RPA & SingerSim \\
\midrule
SepFormer & 0.780 & 0.773 & 0.955 & 0.848 & 0.727 & 0.890 & 0.462 & 0.771 \\
MossFormer & 0.389 & \underline{0.847} & \textbf{0.982} & 0.638 & 0.783 & 0.941 & \underline{0.563} & \underline{0.845} \\
SAM Audio & 0.460 & 0.846 & \underline{0.978} & 0.885 & \textbf{0.865} & \textbf{0.958} & \textbf{0.643} & \textbf{0.909} \\
SR-CorrNet & 1.187 & 0.683 & 0.742 & 1.082 & 0.680 & 0.728 & 0.431 & 0.659 \\
AudioSep & \underline{0.388} & 0.457 & 0.921 & \underline{0.455} & 0.779 & 0.936 & 0.366 & 0.788 \\
ConvTasNet & 1.195 & 0.252 & 0.580 & 1.239 & 0.315 & 0.568 & 0.093 & 0.589 \\
SoloAudio & 1.006 & 0.391 & 0.881 & 1.058 & 0.462 & 0.797 & 0.251 & 0.714 \\
\midrule
\modelname{} & \textbf{0.253} & \textbf{0.854} & \textbf{0.982} & \textbf{0.354} & \underline{0.799} & \underline{0.956} & 0.438 & 0.819 \\
\bottomrule
\end{tabular*}
\end{table}
\FloatBarrier
\subsection{Training Ablation} Speaker-timeline supervision raises SFT count accuracy on every set and lowers PI-WER on noised LibriMix and WildTV, while clean LibriMix PI-WER is nearly unchanged (\cref{tab:ablations}).

\begin{table}[htbp]
\caption{Speaker-timeline ablation: PI-WER $\downarrow$ and count accuracy (Acc.) $\uparrow$. All rows use blind count inference.}
\label{tab:ablations}
\centering\small
\setlength{\tabcolsep}{7pt}\renewcommand{\arraystretch}{1.1}
\begin{tabular}{@{}llrrr@{}}
\toprule
& & \multicolumn{2}{c}{LibriMix} & \\
\cmidrule(lr){3-4}
Recipe & Metric & Clean & Noised & WildTV \\
\midrule
SFT w/o timelines & PI-WER & 0.139 & 0.378 & 0.330 \\
 & Acc. & 0.890 & 0.574 & 0.565 \\
\midrule
Standard SFT & PI-WER & 0.140 & 0.358 & 0.287 \\
 & Acc. & 0.926 & 0.735 & 0.761 \\
\midrule
RL (\modelname{}) & PI-WER & 0.060 & 0.192 & 0.150 \\
 & Acc. & 1.000 & 1.000 & 1.000 \\
\bottomrule
\end{tabular}

\end{table}
\FloatBarrier
\subsection{Judge Validation on Known-Source Mixtures}\label{app:judge:validation}
The judge never observes isolated sources, reference transcripts, or evaluation metrics; it only orders candidates of the same recording. We therefore test whether this ordering agrees with a reference ordering on mixtures whose sources are known. We take 40 LibriMix mixtures (20 clean and 20 noised) and, for each, eight separation outputs randomly drawn from the inference results of intermediate supervised checkpoints. LibriMix sources are clean read speech, so PI-WER reflects most quality differences between separation outputs and serves as the reference ordering. The judge describes each track with Qwen3-Omni, ranks each group twice in text with the candidate order reversed, and applies the order-consistency check. In this test the text ranker runs at temperature 0, and dead and below-anchor candidates receive a common floor score; pairs are compared by the final judge scores.

With Qwen3.6-27B as the text ranker, 36 of the 40 groups pass the order-consistency check. These groups contain 307 output pairs whose PI-WER differs by more than 0.2. The judge prefers one output in 207 of them and agrees with the PI-WER order in 184, or 88.9\% (95\% confidence interval 80.8--95.6\% from 10{,}000 bootstrap resamples of groups). Per group, Kendall's $\tau_b$ between the judge scores and the PI-WER order has a median of 0.40 (interquartile range 0.16--0.57) over the 31 passing groups whose judge scores are not all tied, and it is positive in 29 of them.

\begin{table}[htbp]
\caption{Judge validation on 40 LibriMix groups of eight outputs. Agreement is measured on pairs whose PI-WER differs by more than 0.2, excluding pairs the judge ties.}
\label{tab:judgevalidation}
\centering\small
\begin{tabularx}{\linewidth}{@{}Ycc@{}}
\toprule
Ranking protocol & Groups passing check & Agreement with PI-WER order \\
\midrule
Direct ranking from audio (Qwen3-Omni) & 0/40 & --- \\
Describe, then rank in text (Qwen3.6-27B) & 36/40 & 88.9\% \\
\bottomrule
\end{tabularx}
\end{table}

\FloatBarrier

The same test compared ranking protocols. Ranking all outputs of a group directly from audio with Qwen3-Omni passed the order-consistency check in none of the 40 groups, whereas describing each clip in isolation and ranking the descriptions in text passed in 36 (\cref{tab:judgevalidation}). Together with the rubric-score fluctuations noted in \cref{sec:judge}, this result motivated the describe-then-rank design. The reference ordering reflects transcript accuracy only, so the test does not measure voice purity or audio quality beyond their effect on transcription. Separately, the RL model has lower PI-WER than standard SFT on both LibriMix conditions and on WildTV (\cref{tab:ablations}), although the reward uses no reference transcripts.

\FloatBarrier
\section{Application: Full-Duplex Conversation Processing}\label{sec:duplex}\label{app:duplex}
\begin{figure}[htbp]
\centering
\includegraphics[width=0.9\linewidth]{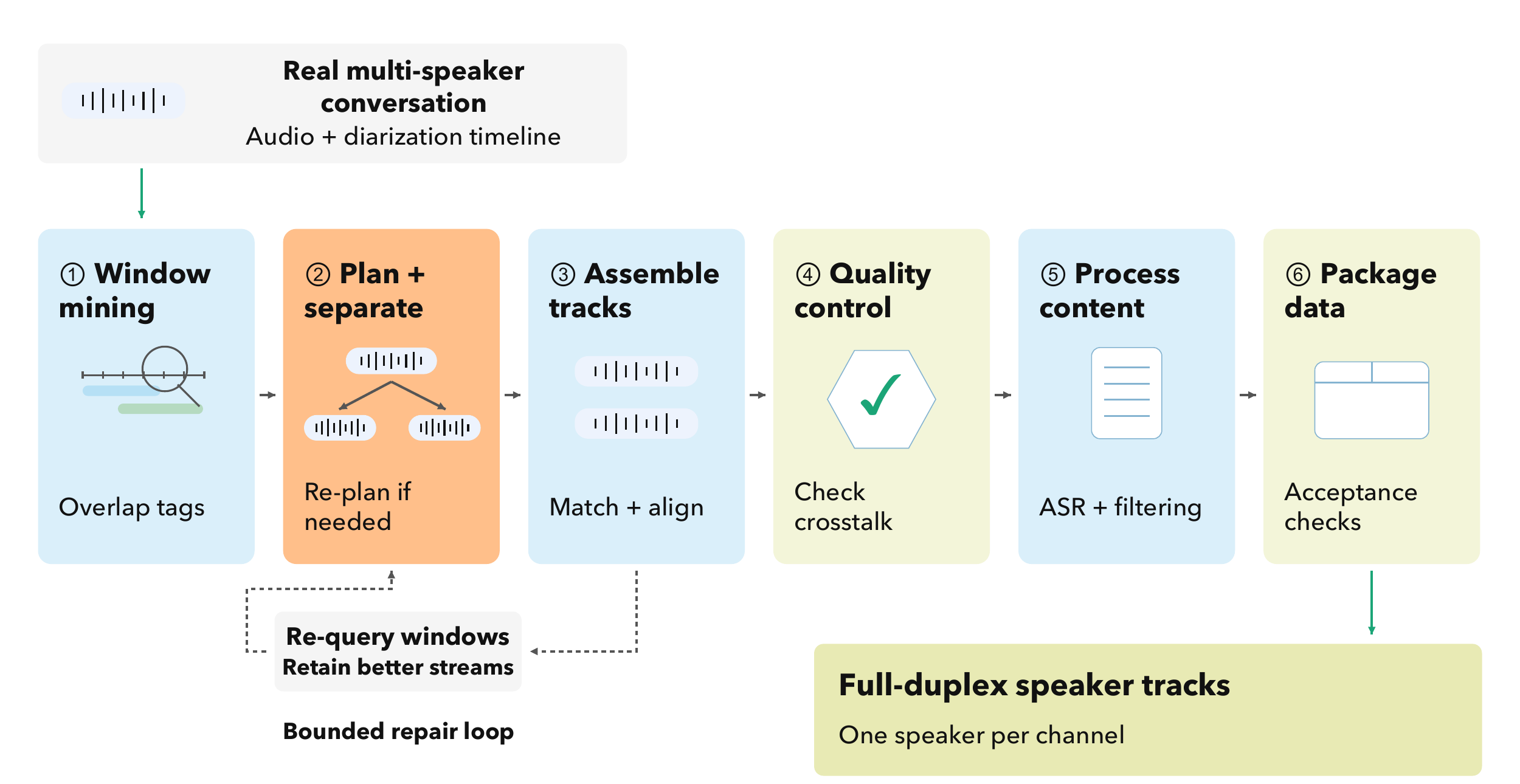}
\caption{Full-duplex conversation processing. Speaker activity guides window mining and adaptive separation. Voice embeddings link outputs across pieces, and alignment places them on a shared recording timeline. The dashed loop re-queries failed windows and retains improved outputs. Separate identity, missing-speech, and content checks precede transcript-aligned packaging.}
\label{fig:duplex}
\end{figure}

Full-duplex conversations require time-aligned speaker tracks that preserve both overlapping speech and short backchannels. We use \modelname{} as the separation component of an offline pipeline that converts diarized recordings into continuous per-speaker audio with aligned transcripts (\cref{fig:duplex}). Speaker timelines determine the active speakers in each inference piece; every piece, including single-speaker regions, passes through the model.

The pipeline converts a conversation recording with $K$ speakers into $K$ continuous speaker tracks with word-aligned transcripts; our examples use $K=2$ and $K=3$. Its input is the recording and a speaker timeline with transcripts from MOSS-Transcribe-Diarize~\citep{moss_transcribe_diarize_2026}. Processing follows four stages: window mining and planning (\cref{app:duplex:planning}), separation and assembly (\cref{app:duplex:assembly}), quality control (\cref{app:duplex:quality}), and acceptance (\cref{app:duplex:acceptance}). \Cref{tab:duplexparams} lists the pipeline parameters. The supplementary material contains audio examples processed by the pipeline: eight two-speaker conversations (four Chinese and four English) and six Chinese three-speaker conversations with three output tracks, each with the input mixture, the separated speaker tracks, and aligned transcripts.

\subsection{Window Mining and Planning}\label{app:duplex:planning}
\paragraph{Window mining.}
The pipeline first selects conversation windows that are long enough to capture turn-taking and reliable enough to annotate. Each window contains exactly $K$ speakers, gives each speaker enough speech to keep short backchannels, including at least 1 s of exclusive speech, and is dense in speech. Boundaries fall in silence gaps and avoid the end of a possibly truncated transcript. Windows with ambiguous multi-speaker labels, near-duplicates of another window from the same recording, or a suspicious single-block speaker annotation are discarded. For $K=2$, each retained window is assigned to the overlap or turn-taking stratum according to whether the timeline contains a sufficiently long speaker intersection; overlap that the timeline misses is therefore not reflected in the strata. Windows with $K=3$ form a single stratum.

\paragraph{Piece planning.}
A conversation window is too long to separate in one pass, so it is split into inference pieces. A dynamic program chooses pieces that cover all annotated speech while minimizing the sum of piece duration times requested stream count, a proxy for the number of generated audio tokens. Each piece requests one stream per speaker active in it, rather than one per speaker in the whole window, and adds context on both sides for later cross-fading. Long monologues are split before planning, and a soft penalty, proportional to the missing duration and the stream count, discourages pieces shorter than the preferred length for their stream count. A piece requests at most $\min(K,3)$ streams.

\subsection{Separation and Assembly}\label{app:duplex:assembly}
\paragraph{Separation and speaker assignment.}
\modelname{} separates every planned piece, including single-speaker pieces. CAM++ embeddings~\citep{wang2023campp} then assign the local streams to the $K$ conversation speakers, and a cross-piece consistency term keeps assignments stable across neighboring pieces. A stream is rejected as mixed when its second-highest speaker similarity is also high; the margin adapts to the similarity between the speaker references, so that speakers with similar voices are not rejected wholesale.

\paragraph{Placement.}
Each assigned stream is placed on the recording time axis relative to its speaker's timeline onset. When its confidence is high enough, energy-envelope alignment refines the offset within a bounded range. Neighboring pieces are cross-faded over their shared context. Timeline masks, slightly expanded but never entering other speakers' intervals, remove residual speech outside the speaker's own activity, and all zeroed regions are filled with a low-level noise floor sampled from the track's quietest frames. Speaker verification over each speaker's full timeline intervals then checks the assembled assignment. For $K=2$, the two tracks are swapped only when both assignments are reversed; for $K\geq3$, the Hungarian algorithm reassigns tracks whenever a better permutation exists. Streams that remain misassigned after this check are discarded.

\paragraph{Repair.}
Pieces that return too few usable streams, and streams rejected during assignment, are re-queried. The repair window keeps the active-speaker set and strictly contains the previous interval, so a replacement cannot open new coverage gaps. For a piece that already had enough streams, a replacement is accepted only if it improves separation, judged by the energy contrast between speaker streams and by envelope agreement in their exclusive regions; when those regions are too short to compare, the existing output is kept. For a piece with missing streams, a replacement is accepted once it provides enough voiced streams. Repair stops at a round limit or when the number of empty slots stops decreasing appreciably, and unrecoverable streams remain explicitly marked as missing.

\subsection{Quality Control}\label{app:duplex:quality}
Each assembled track is checked for three failure modes, each requiring different evidence.

\paragraph{Intrusive speech.}
MOSS-Transcribe-Diarize re-diarizes each track. When it detects more than one speaker, speaker verification localizes the intrusive intervals with CAM++ and ERes2Net~\citep{chen2023eres2net}; a region is kept only when both verifiers accept the target speaker, and scrubbed intervals are filled with the noise floor. An utterance that is mostly scrubbed is removed from the timeline. The track fails the purity check if minority-speaker speech that the verifiers cannot resolve (score gap within 0.10) accumulates beyond the threshold, or if verification cannot cover it. Combining diarization with verification separates true intrusions from diarization over-segmenting a single speaker.

\paragraph{Missing speech.}
Hole detection compares the track with the mixture over the source speaker timeline. The hole fraction is measured over frames in the speaker's exclusive regions where the mixture has speech-level energy.

\paragraph{Mismatched content.}
A track can carry the right voice with the wrong words. Qwen3-ASR-1.7B and Qwen3-ForcedAligner-0.6B~\citep{shi2026qwen3asr} produce the track transcript and word timestamps, and the word count is compared with MOSS-Transcribe-Diarize. Within exclusive-speaker regions, an utterance whose normalized energy envelope correlates poorly with the mixture is marked as mismatched, with an amplitude-domain fallback. Instead of discarding the whole conversation, the pipeline cuts out mismatched utterances by splitting all tracks and the mixture at those utterances into separate samples; pieces shorter than 20 s are dropped, so final samples can be shorter than 60 s. Regions with too little exclusive speech are marked unassessable and reported separately.

\subsection{Acceptance and Output}\label{app:duplex:acceptance}
A conversation is accepted when every speaker has a track and transcript, every track passes the purity check, each speaker's completeness reaches the threshold, the speaker with the least speech contributes enough, and no track fails the hole gate. The hole gate requires both a large hole fraction and a long missing duration, so very short exclusive regions are not rejected on their ratio alone. Completeness is an automatic proxy that can exceed one, so it is used only for coarse filtering. A batch must also stay within the failure-rate limits in \cref{tab:acceptance}, and every delivered track must be non-silent. Listening to stress cases complements these automatic checks.

Each accepted conversation provides one 24 kHz mono track per speaker (and, for two speakers, a stereo version with one speaker per channel), together with speaker intervals, utterances, and word timestamps on the same time axis. Speaker assignments are marked as model-derived. A filtered export applies all gates, while an unfiltered export keeps the hole, purity, and completeness metadata for downstream selection.

\FloatBarrier
\section{Conclusion}\label{sec:conclusion}
We introduced \modelname{}, a generative model that separates individual voices from a mixture without a target-speaker query or a supplied voice count. Its training combines source-audio and speaker-timeline supervision with reinforcement learning on real recordings, where a model judge compares independently obtained track descriptions under repeated candidate orderings.

With supplied voice counts, \modelname{} achieves the lowest transcription error among the compared separators on noisy overlapping speech, speech in noisy environments, and singing vocals. On real-world multi-speaker recordings with unknown voice counts, it also separates voices well, while the singing results show that lyric content, pitch accuracy, and vocal similarity capture different aspects of separation quality. We also integrate the model into an offline pipeline that assembles continuous speaker tracks and aligned transcripts from real conversations. Together, the model and recipe provide a way to learn voice separation from constructed mixtures and natural recordings with incomplete references.

\FloatBarrier
\label{main:end}
\bibliographystyle{unsrtnat}
\bibliography{main}
\clearpage
\appendix
\section{Model and Training Details}\label{app:configuration}
\subsection{Model Configuration}
\begin{table}[htbp]
\caption{Model configuration.}
\label{tab:config}
\centering\small
\begin{tabularx}{\linewidth}{@{}lY@{}}
\toprule
Component & Configuration \\
\midrule
Backbone & Qwen3-4B~\citep{yang2025qwen3} \\
Continuous encoder & Qwen3-Omni AuT; trainable \\
Adaptors & Gated MLPs: $2560\!\to\!8192\!\to\!2560$ for codec embeddings and $2048\!\to\!8192\!\to\!2560$ for AuT features; separate output-code adaptor \\
Local transformer & One layer; hidden size 2560; 32 heads of dimension 80; twelve codebooks with 1024 entries plus padding \\
Codec & MOSS-Audio-Tokenizer-v2; first 12 of 32 residual quantizers; frozen \\
Parameter count & 5,385,239,424 \\
\bottomrule
\end{tabularx}
\end{table}
\FloatBarrier
\subsection{Optimization Settings}\label{app:training:rl}
For the token likelihood ratio $\rho_{it}=\pi_\theta(z_{it}\mid z_{i,<t},x)/\pi_{\mathrm{old}}(z_{it}\mid z_{i,<t},x)$ and advantage $A$, the clipped surrogate is
\begin{align}
 b(\rho,A)&=\min\!\left(\rho A,\clip(\rho,0.8,1.28)A\right),\\
 u(\rho,A)&=\begin{cases}\max(b(\rho,A),3A),&A<0,\\b(\rho,A),&A\geq0.\end{cases}
\end{align}
The policy loss averages the negative surrogate over eligible tokens, with a KL penalty and a codec cross-entropy auxiliary term. Audio likelihood ratios use truncated importance sampling. Centered per-stream feedback augments audio-token advantages as specified in \cref{app:judge:stream}.
\begin{table}[htbp]
\caption{RL configuration for both phases. Only the learning rate and KL coefficient change between phase 1 and phase 2.}
\label{tab:rlsettings}
\centering\small
\begin{tabularx}{\linewidth}{@{}lY@{}}
\toprule
Setting & Value \\
\midrule
Group and batch & $G=8$ candidates/prompt; 48 retained groups/update; 384 sequences \\
Dynamic sampling & Initial 72 attempted groups; cap 400; filter reward std $\leq0.005$ \\
Advantage & Group-centered reward; no standard-deviation normalization \\
Policy clipping & Lower/upper ranges 0.2/0.28; negative-advantage dual clip 3.0~\citep{ye2020mastering} \\
KL and entropy & $k_1$ KL estimator; coefficient 0 in the first phase and 0.001 in the second; no entropy bonus \\
Optimizer & Adam; learning rate $10^{-6}$ in the first phase and $2\times10^{-6}$ in the second; 10 warm-up updates then constant; $(\beta_1,\beta_2)=(0.9,0.98)$; $\epsilon=10^{-8}$; no weight decay \\
Rollout sampling & Temperature 1.0; top-$p=1.0$; maximum response 2048 tokens \\
Audio policy term & Coefficient 1.0; truncated importance sampling enabled \\
Stream feedback & Coefficient 0.5 on centered stream scores \\
Codec auxiliary & Weight 0.05; positive-advantage rollouts only; codebook decay 0.8 \\
\bottomrule
\end{tabularx}
\end{table}
\FloatBarrier
In \cref{eq:rl-loss}, the enabled KL coefficient is $\beta_{\mathrm{KL}}=0.001$ with the $k_1$ estimator, the auxiliary coefficient is $\lambda_{\mathrm{CE}}=0.05$, and centered stream feedback has weight $\lambda_{\mathrm{stream}}=0.5$. The group size is $G=8$ and the spread threshold is $\sigma_{\min}=0.005$. Phase 1 uses $\beta_{\mathrm{KL}}=0$; phase 2 uses the stated coefficient.

\subsection{Output Interface and Exact Prompts}\label{app:model:prompts}
The model retains the Qwen3 vocabulary and chat template, assigning audio roles to existing tokens (\cref{tab:tokens}). At each input-frame position, it sums the projected continuous feature and the embedded mixture codes. At an output-frame request, the local transformer generates twelve codes and the output adaptor embeds them for subsequent backbone steps.

\begin{table}[htbp]
\caption{Vocabulary tokens reused for the audio interface.}
\label{tab:tokens}
\centering\small
\begin{tabularx}{\linewidth}{@{}lY@{}}
\toprule
Token & Role \\
\midrule
\texttt{<|im\_start|>}, \texttt{<|im\_end|>} & Turn start and turn end / stop \\
\texttt{<|vision\_start|>}, \texttt{<|vision\_end|>} & Audio boundaries for input and output \\
\texttt{<|vision\_pad|>} & Input frame, replaced by the mixture embedding \\
\texttt{<|video\_pad|>} & Output frame request, replaced by the generated-code embedding \\
\texttt{<|endoftext|>} & Sequence padding \\
\bottomrule
\end{tabularx}
\end{table}
\FloatBarrier

Teacher forcing supervises all assistant text-channel positions, including indices, counts, delimiters, frame requests, newlines, and the stop token. Audio-code cross-entropy acts only at output-frame positions. Separation targets follow first onset; the timeline task uses the text channel alone.

The two requests below specify the separation and timeline tasks. The fixed Chinese strings are the actual model inputs. $\times T$ denotes repeated frame placeholders, and the response values illustrate the format.
\paragraph{Separation.}
\begin{CJK}{UTF8}{gbsn}
\begin{promptbox}
\small\ttfamily
<|im\_start|>system\\
你是一个人声分离助手。<|im\_end|>\\
<|im\_start|>user\\
请先输出这段混合音频中的说话人个数，再分别分离并输出每一段人声，只输出编号和对应分离音频：<|vision\_start|><|vision\_pad|>{\rmfamily$\times T$}<|vision\_end|><|im\_end|>\\
<|im\_start|>assistant\\
2\\
1<|vision\_start|><|video\_pad|>{\rmfamily$\times T_1$}<|vision\_end|>\\
2<|vision\_start|><|video\_pad|>{\rmfamily$\times T_2$}<|vision\_end|><|im\_end|>
\end{promptbox}
\end{CJK}
The system instruction asks the model to act as a voice separation assistant. The user instruction requests the speaker count followed by indexed separated audio. The mixture immediately follows the colon. Each audio placeholder denotes one frame of twelve codes.

\Needspace{12\baselineskip}
\paragraph{Timeline.}
\begin{CJK}{UTF8}{gbsn}
\begin{promptbox}
\small\ttfamily
<|im\_start|>system\\
你是一个人声分离助手。<|im\_end|>\\
<|im\_start|>user\\
请输出这段混合音频中每个说话人的发声时间段，格式为[开始时间][说话人编号][结束时间]，按开始时间排序，不要输出语音内容：<|vision\_start|><|vision\_pad|>{\rmfamily$\times T$}<|vision\_end|><|im\_end|>\\
<|im\_start|>assistant\\
{[}0.06{]}{[}S01{]}{[}1.22{]}{[}1.44{]}{[}S02{]}{[}3.80{]}{\rmfamily\,$\cdots$\,}<|im\_end|>
\end{promptbox}
\end{CJK}
The user instruction requests speaking intervals in [start][speaker][end] form, sorted by start time, without speech content. Times are seconds to two decimal places. Speaker IDs follow first appearance and can recur across intervals. This response has no count token or audio-frame placeholders.

\FloatBarrier

\FloatBarrier
\clearpage
\section{Judge Settings, Reward Constants, and Prompts}\label{app:judge}
\subsection{Inference Settings}
The perception and ranking settings used by the reward pipeline are listed below. Failure handling is described in \cref{sec:judge:inference}.
\begin{table}[htbp]
\caption{Judge inference settings; tier scores are averaged over two rankings.}
\label{tab:judgeconfig}
\centering\small
\begin{tabularx}{\linewidth}{@{}lY@{}}
\toprule
Component & Setting \\
\midrule
Perception & Qwen3-Omni-30B-A3B-Instruct; greedy; maximum 400 tokens; repetition penalty 1.05 \\
Text ranker & Qwen3.6-27B; thinking disabled; temperature 0.3; top-$p=0.9$; maximum 1024 tokens; repetition penalty 1.05 \\
Candidate presentation & $M=2$ calls; randomly shuffled labels, then reversed order; anonymous mixture anchor included \\
Tier values $(q_A,q_B,q_C,q_D)$ & A/B/C/D $\mapsto$ 1.0/0.7/0.4/0.1 \\
Consistency threshold & $\tau_{\min}=0.3$; accept Kendall's $\tau_b\geq\tau_{\min}$ \\
Group rejection & At most one candidate surviving the signal checks; failed mixture or anchor description; fewer than two fully described surviving candidates; failed ranking call; low rank agreement; both calls indistinguishable; conflicting indistinguishable flags \\
\bottomrule
\end{tabularx}
\end{table}
\FloatBarrier
\subsection{Signal Gates and Reward Shaping}\label{app:judge:constants}
Signal checks operate at 16 kHz with the parameters in \cref{tab:gates}. The number of streams passing the duration, level, and activity tests is $J_{\mathrm{eff}}$. Count penalties relative to the annotation $K$ use this quantity; protocol consistency compares the declared count $\Khat$ with the emitted count $J$.
\begin{table}[htbp]
\caption{Signal-gate parameters for reward shaping.}
\label{tab:gates}
\centering\small
\begin{tabularx}{\linewidth}{@{}YY@{}}
\toprule
Check & Setting \\
\midrule
Stream duration & $d_k>0.2$ s; duration ratio lower bound 0.35, upper bound 1.05 \\
Signal level and activity & RMS $>-45$ dBFS; active fraction $\geq0.05$ \\
Activity frames & 50 ms; active above $-40$ dBFS \\
Log-mel representation & FFT 512; hop 256; 40 mel bands; frame-match tolerance 1.5 \\
Input copy & Frame-match threshold 0.8; per-stream cap 0.1 \\
Duplicate stream & Frame-match threshold 0.8; per-stream cap 0.3 \\
Excess valid streams & Multipliers 1.0/0.7/0.4 for excess 0/1/2; 0 beyond \\
Missing valid streams & Multipliers 1.0/0.5/0.2 for deficit 0/1/2; 0 beyond \\
Declared/emitted count mismatch & Multiplier 0.6 \\
Dead categories & No streams; no valid streams; overlong; zero count penalty; mixture passthrough \\
\bottomrule
\end{tabularx}
\end{table}
\FloatBarrier
Copy and duplicate checks compare log-mel frames, allowing tolerance for codec reconstruction. The configured copy exemption permits a mixture-like output when the declared count is one and the count annotation is at most one. This accommodates single-source reconstruction, while leaving background retention and missed speakers for the subsequent quality assessment.

For the dead-candidate branch of \cref{eq:judge}, severity is $s_i=0$ for passthrough, no streams, or no valid streams; $s_i=0.4$ for overlong output; and $s_i=0.6$ for a zero count penalty. These values put dead candidates below surviving candidates within an accepted group. Rejected groups receive zero reward throughout.
\begin{table}[htbp]
\caption{Reward-mapping constants (\cref{eq:judge}).}
\label{tab:rewardconstants}
\centering\small
\begin{tabularx}{\linewidth}{@{}lrY@{}}
\toprule
Parameter & Value & Role \\
\midrule
$\delta$ & 0.005 & Dead-candidate maximum \\
$\eta$ & 0.25 & Minimum fraction for dead rewards \\
$s_{\max}$ & 0.6 & Maximum category severity \\
$\epsilon_{\mathrm{alive}}$ & 0.006 & Surviving-candidate guard \\
$f_{\mathrm{below}}$ & 0.05 & Below-anchor floor \\
$\beta_{\mathrm{below}}$ & 0.3 & Below-anchor scale \\
\bottomrule
\end{tabularx}
\end{table}
\FloatBarrier

Thus $s_i\in\{0,0.4,0.6\}$. The multiplier $\gamma_i$ is the mean per-stream copy/duplicate cap multiplied by the duration and count step penalties, and by the declared/emitted-count factor (0.6 on mismatch). The guard $\epsilon_{\mathrm{alive}}$ is a constructive safeguard: it exceeds $\delta$, while the unguarded product can fall below $\delta$; for example, $0.1\times0.1\times0.4\times0.6=0.0024$. It was inactive for surviving candidates in the recorded runs. The separate floor $f_{\mathrm{below}}$ is larger, so this mapping is not globally monotone in tier score.

\subsection{Exact Judge Prompts}\label{app:judge:prompts}
We list the exact prompts sent to the judge models. In the ranking prompt, \texttt{\{mix\_report\}} and \texttt{\{reports\}} are replaced by the JSON reports for the mixture and for every candidate track, \texttt{\{ids\}} by the anonymous candidate labels, and \texttt{\{gt\_line\}} by a line stating the annotated voice count.
\Needspace{12\baselineskip}
\paragraph{Track description (\texttt{DESCRIBE\_TRACK\_PROMPT}).}
\begin{quote}\small\ttfamily\raggedright
Listen to this single audio clip. Report ONLY what you actually hear. Answer STRICT JSON, no other text:\\
\{"speakers\_audible": <int>, "speaker\_traits": ["<gender/voice, <=6 words each>", ...], "content": "<what is said, verbatim fragments or brief summary, <=30 words>", "language": "<language(s)>", "artifacts": "none|mild|severe", "mostly\_silence": <bool>, "notes": "<=10 words>"\}\\[0.6\baselineskip]
"speakers\_audible" counts DISTINCT speakers you can hear in this clip (0 if no speech). "artifacts" = distortion, robotic/garbled audio, clicks, unnatural sound.
\end{quote}

\paragraph{Mixture description (\texttt{DESCRIBE\_MIX\_PROMPT}).}
\begin{quote}\small\ttfamily\raggedright
Listen to this audio: it is a MIXTURE that may contain several people speaking, possibly overlapping. Report ONLY what you actually hear. Answer STRICT JSON, no other text:\\
\{"num\_speakers": <int>, "speakers": [\{"traits": "<gender/voice, <=6 words>", "content": "<what this speaker says, <=20 words>"\}, ...], "background": "<=8 words", "notes": "<=10 words"\}
\end{quote}

\paragraph{Text ranking (\texttt{TEXT\_RANK\_PROMPT}).}
\begin{quote}\small\ttfamily\raggedright
You are ranking candidate results of blind speech separation performed on the same input mixture. You cannot hear the audio. An independent listener described the MIXTURE, and separately described every track of every candidate (each track heard IN ISOLATION, without context).\\[0.6\baselineskip]
MIXTURE report:\\
\{mix\_report\}\\
\{gt\_line\}\\
Candidates, labeled \{ids\} (order is arbitrary and meaningless). Each candidate is one attempt to split the mixture into one-speaker tracks:\\
\{reports\}\\[0.6\baselineskip]
Rank ALL candidates into quality tiers A (clearly best) > B > C > D (clearly worst). Priorities, most important first:\\
1. A track with 2+ speakers audible is NOT separated \textemdash{} the more such tracks, the worse. A candidate whose single track sounds like the whole mixture is worst.\\
2. The candidate's tracks together should cover all mixture speakers/content; missing speakers or missing content is bad.\\
3. Two tracks with the same speaker/content = duplicates, bad.\\
4. Track content that does NOT appear in the mixture report = likely fabricated, serious defect.\\
5. Fewer artifacts is better.\\
Ties allowed. Use multiple tiers whenever the reports differ materially.\\[0.6\baselineskip]
Answer STRICT JSON only:\\
\{"indistinguishable": <bool>, "ranking": [\{"id": "C0", "tier": "A"\}, ...one entry per candidate, every id exactly once...], "notes": "<=20 words"\}
\end{quote}

\FloatBarrier
\clearpage
\section{Full-Duplex Pipeline Parameters}\label{app:duplex:parameters}
The parameters below implement the processing stages in \cref{sec:duplex}. Batch limits are configured acceptance thresholds, not observed failure rates.
\begin{table}[htbp]
\caption{Full-duplex pipeline parameters.}
\label{tab:duplexparams}
\centering\footnotesize\renewcommand{\arraystretch}{1.0}
\setlength{\tabcolsep}{4pt}
\begin{tabularx}{\linewidth}{@{}llY@{}}
\toprule
Stage & Parameter & Value \\
\midrule
Window mining & Window length & 60--180 s; $K$ speakers \\
 & Minimum speech per speaker & 0.2 s \\
 & Minimum exclusive speech per speaker & 1 s \\
 & Minimum speech density & 0.45 \\
 & Boundary silence gap / outer padding & $\geq$0.25 s / 0.2 s \\
 & Excluded transcript tail & Final 2 s \\
 & Overlap stratum ($K=2$) & Longest annotated intersection $\geq$0.30 s \\
 & Near-duplicate rejection & Temporal IoU $>$0.30 within a recording \\
 & Single-block rejection & $\leq$2 intervals cover $>$90\% of a span that is $\geq$15 s or $>$20\% of the window \\
\midrule
Piece planning & Maximum piece length / streams & 22 s / $\min(K,3)$ \\
 & Context and speaker-inclusion margin & 0.5 s per side \\
 & Preferred minimum length (1/2/3 streams) & 3/6/8 s; penalty 4.0 per missing second per stream \\
\midrule
Assembly & Cross-piece consistency weight & 2.0 \\
 & Mixed-stream rejection & Highest similarity $\geq$0.45 and second-highest $\geq\max(0.45,\ \text{reference similarity}+0.10)$ \\
 & Alignment refinement & $\pm$6 s; confidence $\geq$0.05 \\
 & Timeline-mask expansion & 0.2 s per side \\
\midrule
Repair & Stopping rule & Round limit, or empty-slot reduction $<$10\% \\
 & Minimum exclusive region for comparison & 0.5 s \\
\midrule
Intrusive speech & Intrusion margin & Other-speaker similarity higher by $\geq$0.10 \\
 & Utterance removal & $\geq$70\% scrubbed \\
 & Purity rejection & Unresolved minority speech (verifier score gap within 0.10) $\geq$2.0 s \\
 & Noise-floor fill (all zeroed regions) & 2nd-percentile speech-frame energy + 3 dB (source frames below + 6 dB) \\
\midrule
Content & Envelope-correlation threshold & 0.30 (amplitude fallback 0.12) \\
 & Minimum exclusive speech for assessment & 0.30 s \\
\midrule
Acceptance & Per-speaker completeness & $\geq$0.80 \\
 & Least-speaking speaker & $\geq$8 s and $\geq$5 utterances \\
 & Hole gate & Fraction $>$0.20 and missing duration $>$1.0 s \\
\bottomrule
\end{tabularx}
\end{table}
\FloatBarrier
\begin{table}[htbp]
\caption{Batch acceptance thresholds. Denominators are specified per row; limits are configured thresholds, not measured rates.}
\label{tab:acceptance}
\centering\footnotesize\renewcommand{\arraystretch}{1.0}
\setlength{\tabcolsep}{4pt}
\begin{tabularx}{\linewidth}{@{}lYYr@{}}
\toprule
Check & Failure event & Denominator & Limit \\
\midrule
Inference & Fewer usable streams than requested & Planned pieces & 3\% \\
Assembly: missing & No assigned stream after repair & Speaker slots per piece & 4\% \\
Assembly: placement & Assigned stream silent in its exclusive region & Speaker slots per piece & 6\% \\
Speech coverage & Track more than 40 dB below mixture during timeline speech & Source-timeline speech duration & 5\% \\
Speaker coverage & Unprocessed speaker with more than 3 s of speech & Evaluated conversation windows & 1\% \\
Content matching & Envelope below 0.30 or fallback discrepancy above 0.12 & Assessable utterances & 8\% \\
Silent output & Track RMS below $-60$ dBFS & Delivered tracks & 0 \\
\bottomrule
\end{tabularx}
\end{table}

\FloatBarrier
\end{document}